# Separated Twins or Just Siblings? A Multiplanet System around an M Dwarf Including a Cool Sub-Neptune

Mallory Harris[1], Diana Dragomir[1], Ismael Mireles[1], Karen A. Collins[2], Solène Ulmer-Moll[3,4], Steve B. Howell[5], Keivan G. Stassun[6], George Zhou[7], Carl Ziegler[8], François Bouchy[3], César Briceño[9], David Charbonneau[2], Kevin I. Collins[10], Gábor Fűrész[11], Natalia M. Guerrero[11,12], Jon M. Jenkins[5], Eric L. N. Jensen[13], Martti H. K. Kristiansen[14], Nicholas Law[15], Monika Lendl[3], Andrew W. Mann[15], Hugh P. Osborn[4,11], Samuel N. Quinn[2], George R. Ricker[11], Richard P. Schwarz[2], Sara Seager[11,16,17], Eric B. Ting[5], Roland Vanderspek[11], David Watanabe[18], and Joshua N. Winn[19]

[1] Department of Physics and Astronomy, University of New Mexico, Albuquerque, NM, USA
[2] Center for Astrophysics | Harvard & Smithsonian, 60 Garden Street, Cambridge, MA 02138, USA
[3] Observatoire de Genève, Université de Genève, Chemin Pegasi 51, CH-1290 Sauverny, Switzerland
[4] Physikalisches Institut, University of Bern, Gesellschaftsstrasse 6, 3012 Bern, Switzerland
[5] NASA Ames Research Center, Moffett Field, CA 94035, USA
[6] Department of Physics and Astronomy, Vanderbilt University, Nashville, TN 37235, USA
[7] Centre for Astrophysics | University of Southern Queensland, Toowoomba, QLD, Australia
[8] Department of Physics, Engineering and Astronomy, Stephen F. Austin State University, 1936 North Street, Nacogdoches, TX 75962, USA
[9] Cerro Tololo Inter-American Observatory, Casilla 603, La Serena, Chile
[10] George Mason University, 4400 University Drive, Fairfax, VA 22030, USA
[11] Department of Physics and Kavli Institute for Astrophysics and Space Research, Massachusetts Institute of Technology, Cambridge, MA 02139, USA
[12] Department of Astronomy, University of Florida, Gainesville, FL 32611, USA
[13] Department of Physics & Astronomy, Swarthmore College, Swarthmore, PA 19081, USA
[14] Brorfelde Observatory, Observator Gyldenkernes Vej 7, DK-4340 Tølløse, Denmark
[15] Department of Physics and Astronomy, The University of North Carolina at Chapel Hill, Chapel Hill, NC 27599-3255, USA
[16] Department of Earth, Atmospheric, and Planetary Sciences, Massachusetts Institute of Technology, Cambridge, MA 02139, USA
[17] Department of Aeronautics and Astronautics, Massachusetts Institute of Technology, Cambridge, MA 02139, USA
[18] Planetary Discoveries, Fredericksburg, VA 22405, USA
[19] Department of Astrophysical Sciences, Princeton University, Princeton, NJ 08544, USA
*Received 2023 March 2; revised 2023 September 27; accepted 2023 October 6; published 2023 December 7*

## Abstract

We report the discovery of two TESS sub-Neptunes orbiting the early M dwarf TOI-904 (TIC 261257684). Both exoplanets, TOI-904 b and c, were initially observed in TESS Sector 12 with twin sizes of $2.426^{+0.163}_{-0.157}$ and $2.167^{+0.130}_{-0.118}$ $R_\oplus$, respectively. Through observations in five additional sectors in the TESS primary mission and the first and second extended missions, the orbital periods of the planets were measured to be $10.887 \pm 0.001$ and $83.999 \pm 0.001$ days, respectively. Reconnaissance radial velocity measurements (taken with EULER/CORALIE and SMARTS/CHIRON) and high-resolution speckle imaging with adaptive optics (obtained from SOAR/HRCAM and Gemini South/ZORRO) show no evidence of an eclipsing binary or a nearby companion, which, together with the low false-positive probabilities calculated with the statistical validation software TRICERATOPS, establishes the planetary nature of these candidates. The outer planet, TOI-904 c, is the longest-period M dwarf exoplanet found by TESS, with an estimated equilibrium temperature of 217 K. As the three other validated planets with comparable host stars and orbital periods were observed by Kepler around much dimmer stars ($J_{mag} > 12$), TOI-904 c, orbiting a brighter star ($J_{mag} = 9.6$), is the coldest M dwarf planet easily accessible for atmospheric follow-up. Future mass measurements and transmission spectroscopy of the similar-sized planets in this system could determine whether they are also similar in density and composition, suggesting a common formation pathway, or whether they have distinct origins.

*Unified Astronomy Thesaurus concepts:* Exoplanet systems (484); Exoplanet detection methods (489); Exoplanet astronomy (486); M dwarf stars (982); Cold Neptunes (2132); Mini Neptunes (1063); Transit photometry (1709)

## 1. Introduction

The past decade of exoplanet exploration has revealed that M dwarf stars typically host multiplanet systems of small, rocky planets with orbital periods of $P < 20$ days (Dressing & Charbonneau 2015; Hardegree-Ullman et al. 2019). As these cool, low-mass stars allow for easier detections and characterizations of exoplanets than larger hosts, whether via deeper transits or more significant radial velocity reflex motion, these short-period planets have been prioritized for in-depth study by the exoplanet community. Yet few planets have been found at greater distances from M dwarf hosts to date. Indeed, the Kepler mission only discovered ~30 M dwarf planets at distances of >0.15 au (periods $P > 25$ days; Muirhead et al. 2012; Barclay et al. 2015; Dressing & Charbonneau 2015; Torres et al. 2015, 2017; Morton et al. 2016; Berger et al. 2018), all of which orbit stars dimmer than $V_{mag} = 15$ and thus are not easily accessible for ground-based observations. By prioritizing M dwarf stars and focusing on stars in the solar neighborhood, the Transiting Exoplanet Survey Satellite (TESS) now has the potential to populate this underexplored region of parameter space and provide targets for future characterization.

TESS is conducting a full-sky survey of nearby, bright stars, 75% of which are M dwarfs (Ricker et al. 2015). Unlike Kepler, which observed one region of the sky for 4 yr, the TESS primary mission (PM) observed both hemispheres by cycling through 26 sectors observed for ∼27 day intervals, and it is continuing to reobserve the hemispheres (as well as the ecliptic) in the same fashion through its extended missions. While this observation strategy makes TESS most sensitive to planets with $P \lesssim 10$ days, there exists the potential to observe planets with longer orbital periods in overlapping sectors near the ecliptic poles. Further, TESS is projected to observe many planets with $P \gtrsim 25$ days as single-transit events (Villanueva et al. 2019) that could be recovered in later extended missions. Since its launch in 2018, TESS has found 28 confirmed planets with periods of >25 days, five of which (Cañas et al. 2020; Rodriguez et al. 2020; Fukui et al. 2022; Mann et al. 2022; Schanche et al. 2022) orbit low-mass stars.

In this Letter, we introduce and statistically validate TESS object of interest (TOI) 904 c, a sub-Neptune with a period of 84 days that was observed to produce a single-transit event in four different sectors of the TESS PM and its first and second extended missions (EM1 and EM2). We also introduce another planet in the same system (TOI-904 b) with a period of 10.87 days. Orbiting an early M dwarf (TIC 261257684, TOI-904) at $T_{\text{equil}} \approx 200$ K, TOI-904 c is the coldest M dwarf planet discovered by TESS to date. Only three other transiting planets have been observed around low-mass stars with similar orbital periods: Kepler-186 f ($P = 129.94$ days; Quintana et al. 2014), Kepler-1229 b ($P = 86.83$ days; Morton et al. 2016), and Kepler-1628 b ($P = 76.38$ days; Morton et al. 2016). Unlike these planets, TOI-904 c orbits an M dwarf bright enough ($J_{\text{mag}} = 9.61$) for future mass measurements and atmospheric characterization. Through these additional observations, this system embodies a (currently) unique opportunity to test theories of planet formation around low-mass stars. Further, by measuring the densities and constraining the composition of both planets in this system, we can determine whether these similarly sized planets also have twin compositions and formation histories or if they evolved through distinct formation pathways while orbiting the same host star.

## 2. Observations

### 2.1. TESS Observations

TESS observed TOI-904 in Sectors 12 and 13 (2019 May 21–2019 July 17) of its PM; Sectors 27, 38, and 39 (2020 July 5–2020 July 30 and 2021 April 29–2021 June 24) of its EM1; and Sector 61 (2023 January 18–2023 February 12) of its EM2. TOI-904 was included in the TESS Candidate Target List (Stassun et al. 2019) and monitored in both the 2 minute postage stamp and longer-cadence (30 minutes in PM, 10 minutes in EM1, 200 s in EM2) full-frame images during both missions (see Figure 1). The transit signatures of TOI-904 b and c were initially detected by the Science Processing Operations Center (SPOC; Jenkins et al. 2016), located at the NASA Ames Research Center, in a transit search of Sector 12 with a noise-compensating matched filter (Jenkins 2002; Jenkins et al. 2010, 2020) on 2019 July 1. This filter originally folded a single transit of TOI-904 c onto the second transit of TOI-904 b. Both the SPOC and the Visual Survey Group (Kristiansen et al. 2022), working in conjunction with the TESS Single Transit Planet Candidate (TSTPC) working group, flagged these transit events and attributed both to a single-planet candidate with an orbital period of $18.35 \pm 0.005$ days. The correct period of TOI-904 b, $10.877 \pm 0.001$ days, was identified by the SPOC via a transit search of Sector 13 conducted on 2019 July 27. The transit signature passed all of the diagnostic tests presented in the resulting data validation report (Twicken et al. 2018) and was fitted with an initial limb-darkened transit model (Li et al. 2019). The difference image centroiding test located the source of the transit signature to within $1.''3 \pm 2.''9$. The TESS Science Office reviewed the diagnostic results and issued an alert for TOI-904.01 on 2019 June 23 (Guerrero et al. 2021), and it was then recognized as a TOI on 2019 July 15 on the TESS data alerts web portal at the Massachusetts Institute of Technology.[20]

The SPOC conducted a subsequent multisector search of Sectors 12, 13, 27, 38, and 39 on 2021 July 25, where they recovered 12 transits of TOI-904 b and the signature of TOI-904 c at 4× the true period (see Appendix B). The SPOC found that the difference imaging centroiding test located the source of the transit signature to within $3.''4 \pm 3.''4$. In 2021 August, a member of the TSTPC working group (Hugh Osborn) found an undetected additional transit of TOI-904 c in SPOC-produced light curves near the end of TESS Sector 39, which, together with the Sector 12 transit, constrained the period of this planet to a discrete set of possible aliases. We proceeded to search all previous TESS observations of this star for additional transits of this outer planet and discovered a third clear transit in TESS Sector 27 from which we were able to derive a unique period of $83.999 \pm 0.001$ days. This planet was registered as a community TOI on 2021 September 1, and the TESS Science Office issued an alert for it on 2022 April 20. We searched the TESS observations of TOI-904 for any additional transit signals and found no evidence of additional planets in this system from the transit method.

To investigate possible false-positive scenarios of the planets that were observed in this system, we visually inspected the background of each individual light curve that included a transit of the outer planet to rule out false positives due to asteroids or other anomalies. We also used the Gaia DR3 (Gaia Collaboration et al. 2023) to investigate the possibility of blended objects in the aperture and whether the renormalized unit weight error (RUWE), a measure of the normalized $\chi^2$ of the Gaia observations to the astrometric single-star fit corrected for color and magnitude dependencies (Lindegren et al. 2018; Pearce et al. 2020), indicated that the host star existed in a binary that could be responsible for the observed transits. With a RUWE metric of ≈0.999, TOI-904 showed no obvious indication of being a stellar binary (Pearce et al. 2020). As our preliminary inspection gave no indication that either planet candidate's transits were false positives, we looked to follow-up observations (as described in the following sections) to further validate this system.

### 2.2. Ground-based Photometry

The TESS pixel scale is ∼21″ pixel$^{-1}$, and photometric apertures typically extend out to roughly 1′, which generally results in multiple stars blending in the TESS aperture. An eclipsing binary in one of the nearby blended stars could thus mimic a transit-like event in the large TESS aperture. With additional ground-based photometric observations, we can attempt to (1) rule out or identify nearby eclipsing binaries (NEBs) as potential sources of the detection in the TESS data, (2) check for the transit-like event on-target using smaller photometric apertures than in the TESS images to confirm that the event is occurring on-target or in a star so close to TOI-904 that it was not detected by Gaia DR3 (which is unlikely, as such a star would be too faint unless it is perfectly aligned with the target star, but which is accounted for in the `triceratops` transit probability calculation in Section 3.2), and (3) refine the TESS ephemeris.

---
[20] https://doi.org/10.17909/t9-wx1n-aw08

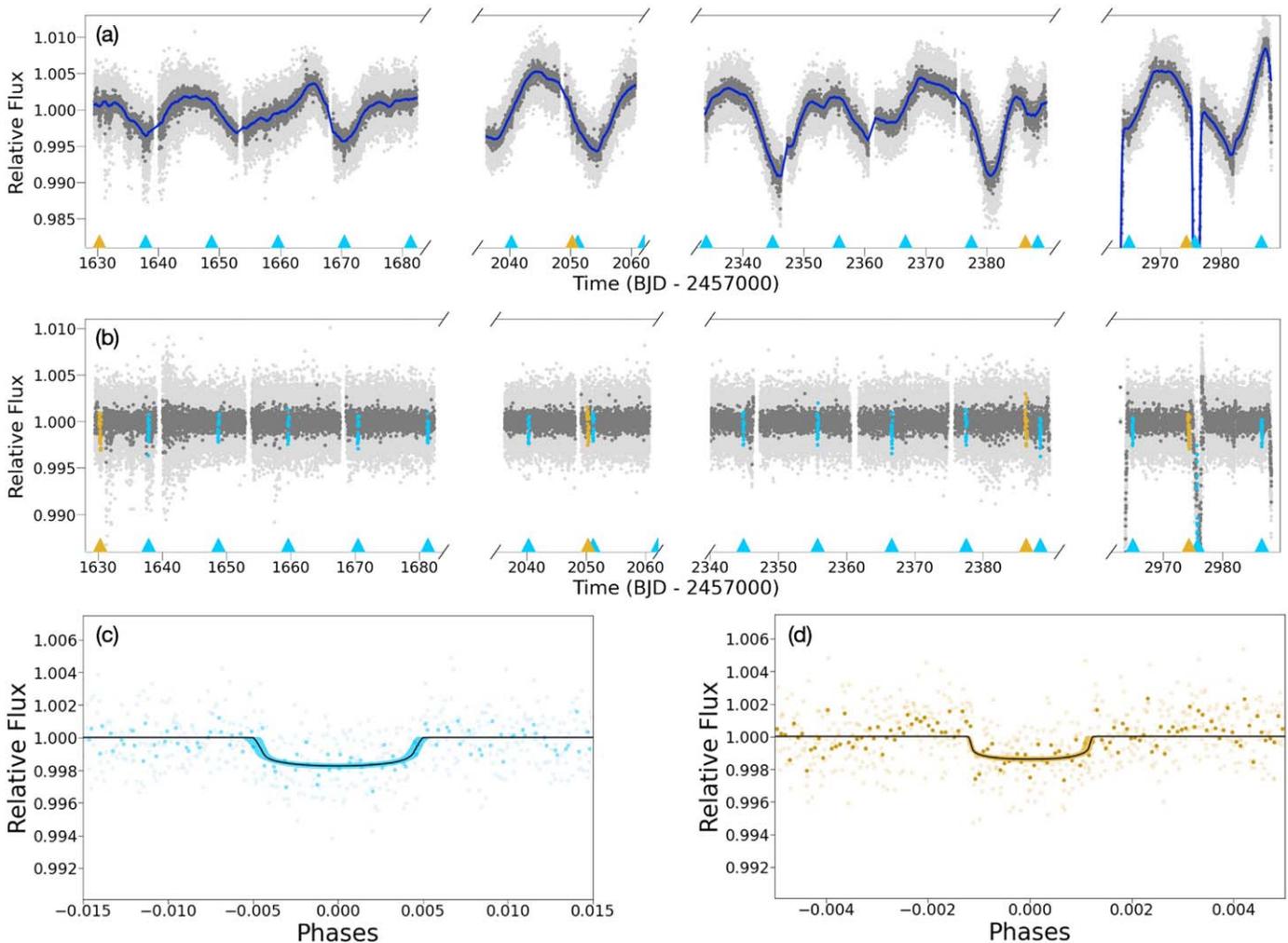

**Figure 1.** TESS observations of TOI-904. Panel (a): raw observations of each TESS sector, with 2 minute data shown in light gray and 10 minute binned data shown in dark gray. We show the Tukey biweight trend calculated with the `wotan` (Hippke et al. 2019) library of the variability in dark blue. Each planet transit is denoted by an arrow at the bottom of the graph, with blue denoting transits of TOI-904 b and gold denoting transits of TOI-904 c. Panel (b): TESS data detrended by the same fit, with each transit highlighted in the same color as the arrows in the previous figure. Panel (c): phase-folded light curves of transits of TOI-904 b, with the transit fit including a shaded region denoting $1\sigma$ uncertainty. Panel (d): phase-folded light curves of TOI-904 c, with the transit fit including a shaded region denoting $1\sigma$ uncertainty.

### 2.2.1. Las Cumbres Observatory

We acquired ground-based transit follow-up photometry of TOI-904 b as part of the TESS Follow-up Observing Program Sub Group 1 (Collins 2019).[21] We used the `TESS Transit Finder`, which is a customized version of the `Tapir` software package (Jensen 2013), to schedule our transit observations and `AstroImageJ` (Collins et al. 2017) to extract differential photometry.

We observed the predicted transit windows of TOI-904 b in the Pan-STARRS $z$-short band from the Las Cumbres Observatory Global Telescope (LCOGT; Brown et al. 2013) 1.0 m network nodes at Siding Spring Observatory and South Africa Astronomical Observatory on UTC 2020 October 15 and 2020 December 19, respectively. The 1 m telescopes are equipped with $4096 \times 4096$ SINISTRO cameras having an image scale of $0.''389$ pixel$^{-1}$, resulting in a $26' \times 26'$ field of view. The images were calibrated by the standard LCOGT `BANZAI` pipeline (McCully et al. 2018).

We used circular photometric apertures of radius $1.''9$ and $5.''8$ to check for possible NEBs that could be contaminating the SPOC photometric apertures, which generally extend $\sim 1'$ from the target star. To account for possible contamination from the wings of neighboring star point-spread functions, we searched for NEBs in all known Gaia EDR3 and TESS Input Catalog (TIC) version 8 nearby stars out to $2.'5$ from TOI-904 that are possibly bright enough in the TESS band to produce the TESS detection (assuming a 100% eclipse and 100% contamination of the TESS aperture). To attempt to account for possible delta-magnitude differences between the TESS band and the follow-up Pan-STARRS z-short band, we checked stars that are an extra 0.5 mag fainter in the TESS band than needed. We consider a star cleared of an NEB if the rms of its 10 minute binned light curve is more than a factor of 5 smaller than the adjusted expected NEB depth in the star (adjusted to allow for the potential TESS-band delta-magnitude difference). We then visually inspect each neighboring star's light curve to ensure that there is no obvious eclipse-like signal. The NEB-checked light-curve data are available at ExoFOP-TESS.[22] We rule out an NEB as the source of the TOI-904 b signal in the TESS data.

---
[21] https://tess.mit.edu/followup

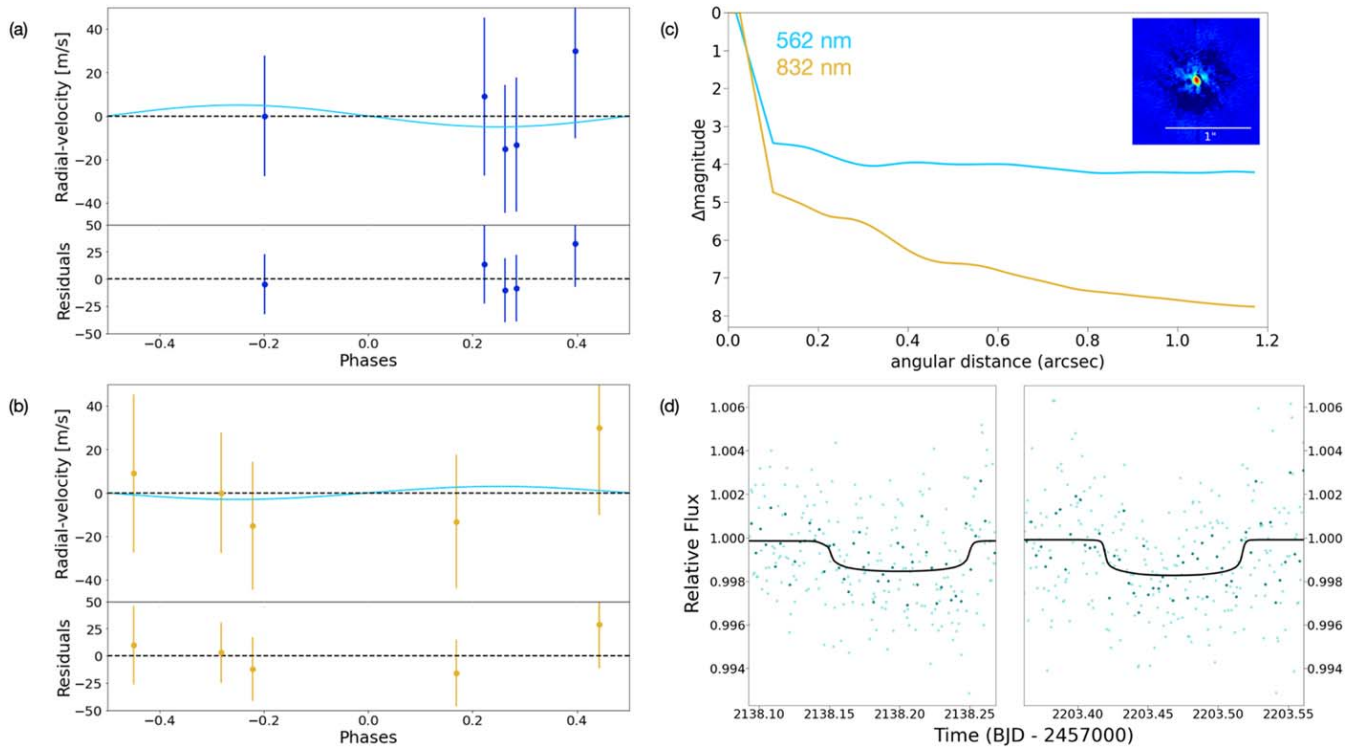

**Figure 2.** Follow-up observations of the planetary system around TOI-904. Left: CORALIE observations phase-folded on the periods of both TOI-904 b (panel (a), dark blue) and TOI-904 c (panel (b), gold). The best-fit radial velocity trends for both phase-folded data sets are shown in light blue. Neither figure shows evidence of a stellar or brown dwarf mass companion. Panel (c): high-resolution imaging taken at 562 and 832 nm (blue and gold, respectively). There is no clear flux from another object between $0''\!.1$ and $1''\!.2$ down to 4 and 8 mag differences, respectively, and thus no evidence of a blending or binary star. Panel (d): ground-based photometric observations of TOI-904 b taken with the LCOGT Siding Spring Observatory (panel (d), left) and the South Africa Astronomical Observatory (panel (d), right). Observations were taken in 36 s intervals (light blue) and binned to 5 minutes (dark blue). The best-fit transit model is overlaid in black, showing that both observations recover full transits of the planet at a comparable depth to the TESS observations.

Photometry of TOI-904 was extracted using circular apertures with a radius of $5''\!.8$, which exclude flux from the nearest known Gaia EDR3 and TIC neighbor (TIC 724109412), $25''\!.4$ southwest. We detect the transit event within the TOI-904 photometric apertures in the two z-short-band light curves and include the data in the analyses of this work (see Figure 2).

We attempted to observe an additional transit of TOI-904 c on UTC 2022 February 27 with the South Africa Astronomical Observatory, resulting in a tentative partial transit detection. This detection, however, lacked a sufficient out-of-transit baseline to be conclusive and is not included in the analyses of this work.

### 2.3. Spectroscopic Follow-up

Brown dwarf or grazing stellar binaries are frequent sources of false-positive transit signals. While few instruments are capable of confirming the exoplanet nature of transits by measuring the masses of small extrasolar planets, many more are equipped to detect the Doppler signal caused by a stellar or brown dwarf mass companion. Reconnaissance radial velocity measurements of the host star are thus able to rule out a false positive due to a bound stellar (or brown dwarf) binary while also providing key observations to refine the spectroscopic parameters of the host star.

#### 2.3.1. SMARTS/CHIRON Spectroscopy

We obtained four observations of TOI-904 with the CHIRON facility on the SMARTS 1.5 m telescope at Cerro Tololo Inter-American Observatory, Chile (Tokovinin et al. 2013). CHIRON is a high-resolution spectrograph with a resolving power of $R = 80,000$ over a wavelength range of 4100–8700 Å with "slicer" mode observations. The spectra were extracted via the standard pipeline as in Paredes et al. (2021).

We attempted to derive a line broadening velocity from the CHIRON observations. We performed a least-squares deconvolution between each spectrum and a nonrotating synthetic spectrum generated via the ATLAS9 model atmospheres grid (Castelli & Kurucz 2004) at the stellar effective temperature and surface gravity of our target star. We then modeled the line broadening profile via a combination of the rotational, radial–tangential macroturbulent, and instrumental broadening kernels (as in Gray & Corbally 1994). We find the rotational broadening component to have a width of <2 km s$^{-1}$ (1$\sigma$) when the additional broadening terms are considered, consistent with expectations from the photometric modulation-derived rotation period of the target star.

---

[22] https://exofop.ipac.caltech.edu/tess/target.php?id=261257684

### 2.3.2. EULER/CORALIE Spectroscopy

TOI-904 was observed by the Swiss 1.2 m Euler telescope with the CORALIE instrument installed at its Nasmyth focus. CORALIE is a fiber-fed high-resolution spectrograph with a spectral resolution of 60,000 (Queloz et al. 2001). CORALIE has a 3 pixel sampling per resolution element. Five spectra of TOI-904 were taken between 2020 February 2 and April 14 (see Figure 2). The observations have an exposure time of 2400 s and reach a signal-to-noise ratio varying between 11 and 16. The observations are targeting the extreme phases of the estimated radial velocity signal (see Figures 2(a) and (b)). The radial velocity is extracted by cross-correlating an M2 stellar mask with each spectrum (Pepe et al. 2002). The radial velocity data have an rms of 15 m s$^{-1}$ over the time span of the observations.

### 2.4. High Angular Resolution Imaging

Close stellar companions (bound or line of sight) can confound exoplanet discoveries in a number of ways. The detected transit signal might be a false positive due to a background eclipsing binary, and even real planet discoveries will yield incorrect stellar and exoplanet parameters if a close companion exists and is unaccounted for (Furlan & Howell 2017, 2020). Additionally, the presence of a bound companion star leads to the nondetection of small planets residing within the same exoplanetary system (Lester et al. 2021).

#### 2.4.1. SOAR/HRCAM High-resolution Imaging

We searched for stellar companions to TOI-904 with speckle imaging with the 4.1 m Southern Astrophysical Research (SOAR) telescope (Tokovinin 2018) on 2022 January 7 UT, observing in the Cousins I band, a similar visible bandpass as TESS. This observation was sensitive to a 5.4 mag fainter star at an angular distance of 1″ from the target. No nearby stars were detected within 3″ of TOI-904 in the SOAR observations.

#### 2.4.2. Gemini South/ZORRO High-resolution Imaging

TOI-904 was observed on 2021 October 21 UT and 2022 January 13 UT using the ZORRO speckle instrument on the Gemini South 8 m telescope (Scott et al. 2021; Howell & Furlan 2022). ZORRO provides simultaneous speckle imaging in two bands (562 and 832 nm) with output data products including a reconstructed image with robust contrast limits on companion detections. While both observations had consistent results that TOI-904 is a single star to within the angular and contrast levels achieved, the 2022 January observation had better seeing, which led to deeper contrast levels. Seven sets of 1000 × 0.06 s images were obtained and processed in our standard reduction pipeline (Howell et al. 2011). Figure 2(c) shows our final contrast curves and the 832 nm reconstructed speckle image. We find that TOI-904 is a single star with no companion brighter than 5–8 mag below that of the target star from the 8 m telescope diffraction limit (20 mas) out to 1″2. At the distance of TOI-904 ($d = 46$ pc), these angular limits correspond to spatial limits of 0.9–55 au.

## 3. Analysis

### 3.1. Host Star Parameters

As an independent determination of the basic stellar parameters from those in the TIC (Stassun et al. 2019), we performed an analysis of the broadband spectral energy distribution (SED) of the star together with the Gaia EDR3 (Gaia Collaboration et al. 2016, 2018) parallax (with no systematic offset applied; see, e.g., Stassun & Torres 2021) in order to determine an empirical measurement of the stellar radius, following the procedures described in Stassun & Torres (2016) and Stassun et al. (2017, 2018). We pulled the $JHK_S$ magnitudes from 2MASS (Skrutskie et al. 2006), the W1–W4 magnitudes from WISE (Wright et al. 2010), the $G_{BP}G_{RP}$ magnitudes from Gaia (Riello et al. 2021), and the near-UV (NUV) flux from GALEX (Martin et al. 2003, 2005). Together, the available photometry spans the stellar SED over the wavelength range 0.2–22 $\mu$m.

We performed a fit using NextGen stellar atmosphere models, with the effective temperature ($T_{\rm eff}$) and metallicity ([Fe/H]) adopted from the spectroscopic analysis using the `SpecMatch-Emp` software tool (Yee et al. 2017) on the CORALIE spectra (the surface gravity, log $g$, has very little influence on the broadband SED). We limited the extinction, $A_V$, to the full line-of-sight value from the Galactic dust maps of Schlegel et al. (1998). The resulting fit has a reduced $\chi^2$ of 2.3 (not including the NUV flux, which suggests the presence of chromospheric activity) with best-fit $A_V = 0.03 \pm 0.03$. Integrating the model SED gives the bolometric flux at Earth, $F_{\rm bol} = 7.62 \pm 0.45 \times 10^{-10}$ erg s$^{-1}$ cm$^{-2}$. Taking the $F_{\rm bol}$ together with the Gaia parallax gives the stellar radius, $R_\star = 0.527 \pm 0.021$ $R_\odot$. The stellar mass can also be estimated via the empirical $M_K$-based relations of Mann et al. (2019), giving $M_\star = 0.557 \pm 0.028$ $M_\odot$.

Finally, we can use the star's rotation period, $P_{\rm rot}$, to estimate its age via empirical gyrochronology relations. We use the full TESS light curve and a Lomb–Scargle periodogram as implemented in `astropy` (Press & Rybicki 1989) to obtain a rotation period of 8.55 days with a false-alarm probability on the order of $10^{-181}$. However, similar spot coverage on opposite hemispheres can cause aliasing at half the true rotation period, meaning that the real rotation period could be 17.1 days. If, in fact, the period is 8.55 days, using empirical relations for M dwarfs from Engle & Guinan (2018) suggests that the stellar age is approximately 1 Gyr. This relatively young age would be consistent with the chromospheric activity suggested by the NUV flux in the SED; however, our current constraints on $v \sin i$ from radial velocity observations of this target favor the 17.1 day period. Precise characterization of the star's $v \sin i$ and other activity indicators, such as log $R'_{\rm HK}$, is needed to better constrain the age of the host star.

These and additional stellar parameters are shown in Table 1.

### 3.2. Planet Validation

In this section, we discuss and rule out false-positive scenarios that could explain the transit signals we have detected

Table 1
Table of TOI-904 Stellar Parameters and the Fitted and Derived Parameters of TOI-904 b and c

| Stellar Parameters | | Planet Parameters | | |
|---|---|---|---|---|
| Catalog Data[a] | | Model Properties | | |
| TIC ID | 261257684 | Fixed Initial Parameters | | |
| TOI | 904 | $q_{1,\mathrm{TESS}}$[d] | | 0.336 |
| R.A. | 05:57:29.11 | $q_{2,\mathrm{TESS}}$[d] | | 0.208 |
| decl. | −83:07:47.02 | $q_{1,zs}$[d] | | 0.281 |
| pmRA (mas yr$^{-1}$) | −28.941 ± 0.032 | $q_{2,zs}$[d] | | 0.192 |
| pmDec (mas yr$^{-1}$) | 110.858 ± 0.033 | Eccentricity | | 0.0 |
| Parallax (mas) | 21.697 ± 0.016 | $\omega$ (deg) | | 90.0 |
| Distance (pc) | 46.089 ± 0.035 | Dilution[a] | | 0.994 |
| Photometric Properties[a] | | Modeled Parameters (SPOC individual fits)[e] | | |
| | | | Planet b | Planet c |
| TESS mag | 10.846 ± 0.007 | T0 (BJD-TDB) | $2366.621^{+0.001}_{-0.002}$ | $2386.349^{+0.003}_{-0.004}$ |
| Gaia mag | 11.8559 ± 0.0004 | Period (days) | $10.8772^{+0.0003}_{-0.0003}$ | $83.9997^{+0.0006}_{-0.0007}$ |
| V mag | 12.588 ± 0.069 | $R_p/R_*$ | $0.039^{+0.001}_{-0.001}$ | $0.038^{+0.001}_{-0.001}$ |
| J mag | 9.607 ± 0.022 | $\rho_*$ (cgs) | $1.878^{+2.710}_{-0.960}$ | $5.545^{+1.304}_{-2.535}$ |
| K mag | 8.766 ± 0.021 | i (deg) | $88.19^{+1.26}_{-0.876}$ | $89.83^{+0.13}_{-0.20}$ |
| Stellar Properties | | Derived Properties | | |
| $M_*$[f] ($M_\odot$) | 0.557 ± 0.028 | $R_p$ ($R_\oplus$) | $2.426^{+0.163}_{-0.152}$ | $2.167^{+0.130}_{-0.118}$ |
| $R_*$[f] ($R_\odot$) | 0.527 ± 0.021 | $R_*/a$ | $0.043^{+0.015}_{-0.009}$ | $0.0078^{+0.0005}_{-0.0014}$ |
| $L_*$[f] ($L_\odot$) | 0.051 ± 0.012 | a (au) | $0.056^{+0.019}_{-0.012}$ | $0.312^{+0.023}_{-0.058}$ |
| $\rho_*$[e] (cgs) | 5.360 ± 0.296 | b | $0.714^{+0.124}_{-0.422}$ | $0.373^{+0.288}_{-0.280}$ |
| $T_{\mathrm{eff}}$[b] (K) | 3770.2 ± 70.0 | $t_{\mathrm{duration}}$ (hr) | $3.65^{+1.27}_{-0.78}$ | $5.04^{+0.37}_{-0.93}$ |
| [Fe/H][b] (dex) | 0.022 ± 0.090 | $T_{\mathrm{eq}}$ (K) | 429.43 ± 32.61 | 217.26 ± 9.10 |
| $v \sin i$[c] (km s$^{-1}$) | <2 | | | |
| Age[f] (Gyr) | 1.5 ± 0.2 | | | |
| | (or 0.8 ± 0.1) | | | |

**Notes.**
[a] Taken from TIC (Stassun et al. 2019).
[b] Derived from CORALIE observations (Section 2.3.2) using SpecMatch-Emp (Yee et al. 2017).
[c] Derived from CHIRON observations discussed in Section 2.3.1
[d] Calculated using LDTK (Husser et al. 2013; Parviainen & Aigrain 2015).
[e] Calculated using juliet (Espinoza et al. 2019) on SPOC (Jenkins et al. 2016) observations.
[f] Section 3.1 analysis.

---

in our observations of TOI-904. Our radial velocity observations taken with the CORALIE instrument sampled the expected times of radial velocity maxima and minima for both planets and detected no signal indicating a stellar or brown dwarf mass companion to the host star. Using the radvel package (Fulton et al. 2018) under the assumption of circular orbits, we obtain 3$\sigma$ mass limits of 238.85 and 350.34 $M_\oplus$ for TOI-904 b and c, respectively, which precludes a stellar mass object at these orbital periods. The high angular resolution images of TOI-904 taken with SOAR and ZORRO show no nearby stars brighter than 5–8 mag less than the host star to distances as close as 0.9 au, from which we can rule out the possibilities that the transits are located on a nearby star or that a nearby star is diluting our observations. Together, the reconnaissance radial velocities and imaging observations lead us to rule out the possibility that a blended eclipsing binary could be causing either of the observed transiting planet candidates.

Our ground-based photometric observations of TOI-904 b further allow us to eliminate the possibility that the TESS observations of TOI-904 b occurred on another star in the TESS pixel. Both observed transits had a strong significance of detection of 11$\sigma$ (see Figure 2(d)). Although we have yet to recover a complete transit of TOI-904 c from ground-based instruments, statistical analyses from the Kepler survey (Lissauer et al. 2012; Rowe et al. 2014; Morton et al. 2016) and recent calculations based on TESS observations from Guerrero et al. (2021) indicate that multiplanet transiting systems are nearly always true planets, especially those with planets smaller that 6 $R_\oplus$, adding validity to the planetary nature of TOI-904 c.

Finally, we used the triceratops (Giacalone & Dressing 2020) software library to statistically validate both planet candidates. triceratops begins by using the Mikulski Archive for Space Telescope (MAST) module of astroquery (Ginsburg et al. 2019) to obtain the TIC properties for each star within 10 pixels of the target star. Using the TESS magnitudes, each neighboring star is considered for its flux contribution in the target pixel and as a possible source of the transit signal. The tool then creates models of transiting planet and eclipsing binary light curves that are used to calculate the probability of each scenario using a Bayesian framework. These probabilities are then used to determine whether or not a planet candidate can be classified as "validated" (has a false-positive probability, FPP, of <1.5% and nearby FPP of <0.1%). Previously, in their work statistically validating

Table 2
Parameters Derived from the `juliet`, `emcee`, and Joint `juliet` Planet Fits for Each Pipeline's Data Products Used in This Study and the LCO Observations of TOI-904 b

| | Pipelines | Planet b | | | | Planet c | | | |
| --- | --- | --- | --- | --- | --- | --- | --- | --- | --- |
| | | $R_p/R_\star$ | $b$ | $i$ (deg) | $R_p$ ($R_\oplus$) | $R_p/R_\star$ | $b$ | $i$ (deg) | $R_p$ ($R_\oplus$) |
| `juliet` | SPOC | $0.042^{+0.002}_{-0.002}$ | $0.71^{+0.12}_{-0.42}$ | $88.19^{+1.26}_{-0.88}$ | $2.42^{+0.16}_{-0.15}$ | $0.037^{+0.002}_{-0.001}$ | $0.37^{+0.29}_{-0.28}$ | $89.83^{+0.13}_{-0.20}$ | $2.17^{+0.13}_{-0.12}$ |
| | TESS-SPOC | $0.042^{+0.002}_{-0.002}$ | $0.79^{+0.10}_{-0.24}$ | $87.73^{+1.11}_{-1.17}$ | $2.46^{+0.18}_{-0.17}$ | $0.037^{+0.002}_{-0.002}$ | $0.42^{+0.29}_{-0.31}$ | $89.80^{+0.15}_{-0.23}$ | $2.20^{+0.27}_{-0.18}$ |
| | QLP | $0.037^{+0.003}_{-0.002}$ | $0.49^{+0.35}_{-0.37}$ | $88.94^{+0.83}_{-1.89}$ | $2.15^{+0.20}_{-0.12}$ | $0.036^{+0.002}_{-0.002}$ | $0.57^{+0.30}_{-0.36}$ | $89.70^{+0.21}_{-0.43}$ | $2.07^{+0.21}_{-0.15}$ |
| | `eleanor` | $0.038^{+0.003}_{-0.002}$ | $0.55^{+0.31}_{-0.34}$ | $88.84^{+0.78}_{-1.77}$ | $2.23^{+0.22}_{-0.14}$ | $0.042^{+0.005}_{-0.003}$ | $0.57^{+0.32}_{-0.22}$ | $89.72^{+0.12}_{-1.50}$ | $2.42^{+0.33}_{-0.20}$ |
| | LCO (10/15/2020) | $0.036^{+0.003}_{-0.003}$ | $0.47^{+0.28}_{-0.30}$ | $89.11^{+0.61}_{-0.97}$ | $2.10^{+0.20}_{-0.19}$ | | | | |
| | LCO (12/19/2020) | $0.039^{+0.003}_{-0.003}$ | $0.36^{+0.23}_{-0.25}$ | $89.38^{+0.44}_{-0.52}$ | $2.23^{+0.18}_{-0.19}$ | | | | |
| `juliet` two-planet fit | SPOC | $0.040^{+0.001}_{-0.001}$ | $0.24^{+0.25}_{-0.17}$ | $89.55^{+0.32}_{-0.56}$ | $2.29^{+0.10}_{-0.10}$ | $0.038^{+0.001}_{-0.001}$ | $0.485^{+0.13}_{-0.08}$ | $89.77^{+0.04}_{-0.08}$ | $2.19^{+0.11}_{-0.11}$ |
| | TESS-SPOC | $0.039^{+0.001}_{-0.001}$ | $0.24^{+0.25}_{-0.17}$ | $89.55^{+0.32}_{-0.56}$ | $2.29^{+0.10}_{-0.11}$ | $0.035^{+0.002}_{-0.001}$ | $0.39^{+0.19}_{-0.19}$ | $89.82^{+0.09}_{-0.11}$[ | $2.03^{+0.12}_{-0.12}$ |
| | QLP | $0.038^{+0.001}_{-0.001}$ | $0.21^{+0.25}_{-0.15}$ | $89.61^{+0.28}_{-0.55}$ | $2.19^{+0.11}_{-0.11}$ | $0.034^{+0.002}_{-0.002}$ | $0.44^{+0.19}_{-0.16}$ | $89.79^{+0.08}_{-0.10}$ | $1.96^{+0.14}_{-0.13}$ |
| | `eleanor` | $0.039^{+0.003}_{-0.002}$ | $0.66^{+0.15}_{-0.24}$ | $88.43^{+0.76}_{-0.90}$ | $2.25^{+0.17}_{-0.14}$ | $0.043^{+0.004}_{-0.003}$ | $0.76^{+0.11}_{-0.18}$ | $89.54^{+0.16}_{-0.22}$ | $2.50^{+0.23}_{-0.20}$ |
| `emcee` | SPOC | $0.041^{+0.003}_{-0.002}$ | $0.65^{+0.18}_{-0.43}$ | $88.48^{+1.11}_{-1.11}$ | $2.38^{+0.20}_{-0.15}$ | $0.037^{+0.002}_{-0.001}$ | $0.35^{+0.26}_{-0.24}$ | $89.84^{+0.11}_{-0.17}$ | $2.14^{+0.19}_{-0.10}$ |
| | TESS-SPOC | $0.040^{+0.002}_{-0.001}$ | $0.46^{+0.31}_{-0.31}$ | $89.09^{+0.65}_{-1.19}$ | $2.34^{+0.17}_{-0.12}$ | $0.037^{+0.002}_{-0.002}$ | $0.43^{+0.30}_{-0.29}$ | $89.81^{+0.13}_{-0.24}$ | $2.14^{+0.15}_{-0.13}$ |
| | QLP | $0.039^{+0.003}_{-0.002}$ | $0.47^{+0.31}_{-0.32}$ | $89.02^{+0.69}_{-1.31}$ | $2.25^{+0.17}_{-0.13}$ | $0.036^{+0.003}_{-0.002}$ | $0.48^{+0.34}_{-0.33}$ | $89.77^{+0.16}_{-0.36}$ | $2.07^{+0.19}_{-0.15}$ |
| | `eleanor` | $0.040^{+0.002}_{-0.002}$ | $0.49^{+0.31}_{-0.34}$ | $88.96^{+0.74}_{-1.44}$ | $2.32^{+0.19}_{-0.16}$ | $0.043^{+0.003}_{-0.002}$ | $0.48^{+0.31}_{-0.32}$ | $89.76^{+0.17}_{-0.33}$ | $2.47^{+0.20}_{-0.16}$ |
| | LCO transits (joint fit) | $0.037^{+0.003}_{-0.004}$ | $0.46^{+0.31}_{-0.31}$ | $89.16^{+0.59}_{-1.21}$ | $2.14^{+0.19}_{-0.25}$ | | | | |

**Note.** The fits created from the `eleanor` light curves generally show a deeper transit depth for the outer planet, TOI-904 c. The QLP light curves presented shallower transits but within an uncertainty of the SPOC and TESS-SPOC observations, as did the fits made from `eleanor` observations of the inner planet. The LCO observations show the inner planet to be both smaller than expected and very comparable with the TESS-found sizes of the outer planet. Ground-based observations of TOI-904 c are needed to further elucidate the difference in the two planets' radii.

several hundred TOIs, Giacalone & Dressing (2020) found the FPP (defined as the summed probability of all scenarios in which a planet-sized object is located on the given host star subtracted from unity) for TOI-904 b to be 3% and designated the planet as a "likely planet." By doing a joint fit using the new Gemini South/ZORRO observations discussed in Section 2.4 and incorporating all TESS transit detections of both planets, we find that for TOI-904 b, the $3\sigma$ upper limit on the FPP = 0.049%. Doing the same calculation for TOI-904 c, we calculate the $3\sigma$ upper limit to be 0.0011%, allowing us to statistically validate both planets with >99% certainty. With the added validity from the planet multiplicity of this system (Lissauer et al. 2012), we can confidently state that both objects observed around this star are planets.

### 3.3. Fitting Planet Parameters

We conducted our analysis on the raw 2 minute simple aperture photometry light curves produced by the SPOC pipeline (Twicken et al. 2010; Morris et al. 2020). We obtained all available data using the MAST portal. We removed long-timescale correlated noise (caused by stellar variability) by applying the `wotan` software library's Tukey biweight algorithm (Hippke et al. 2019) to each sector of TESS observations after masking known transit signals. To conduct a transit analysis of both signals in each light curve, we use the `juliet` software library (Espinoza et al. 2019) built on the `batman` (Kreidberg 2015) transit modeling software and the `dynesty` (Speagle 2020) nested sampling algorithm for calculating Bayesian posteriors and evidences. We used the LDTK (Husser et al. 2013; Parviainen & Aigrain 2015) software to calculate the limb-darkening parameters in the TESS passband based on the stellar properties given in Table 1. We assumed linear ephemerides, circular orbits, and quadratic limb-darkening while fitting the following parameters: $R_p/R_*$, $b$, $T_0$, $P$, and stellar density $\rho_*$. We performed both a joint multiplanet fit and individual fits for each planet (in which we masked out all transits of the other planet). We also did an independent check of the `juliet` individual planet fit analyses using the `batman` modeling package with the `emcee` (Foreman-Mackey et al. 2021) python package. The `emcee` software was used to perform an invariant Markov Chain Monte Carlo sampling, initializing 100 walkers and having each fit take 10,000 steps with 8000 burn-in steps. Using the same limb-darkening values, we fitted the parameters $T_0$, $P$, $R_p/R_*$, $\cos i$, and $a/R_*$. Our single-planet analysis was consistent with the results from the `juliet` analysis (see Appendix A, Table 2).

As an additional cross-check, we repeated this analysis for light curves produced by different pipelines to check whether the transit signals were unaffected by different methods of aperture selection and background subtraction. We repeated our analysis for the 30 minute light curves created by the TESS-SPOC, QLP, and `eleanor` (Feinstein et al. 2019) pipelines (see Appendix A). As with the 2 minute observations, we used the raw un-detrended data before repeating the analysis detailed above using the `emcee` and `juliet` single and multiplanet fits. We found that the transit fits created from the `eleanor`, QLP, and 2 and 30 minute SPOC data were in agreement within $1\sigma$.

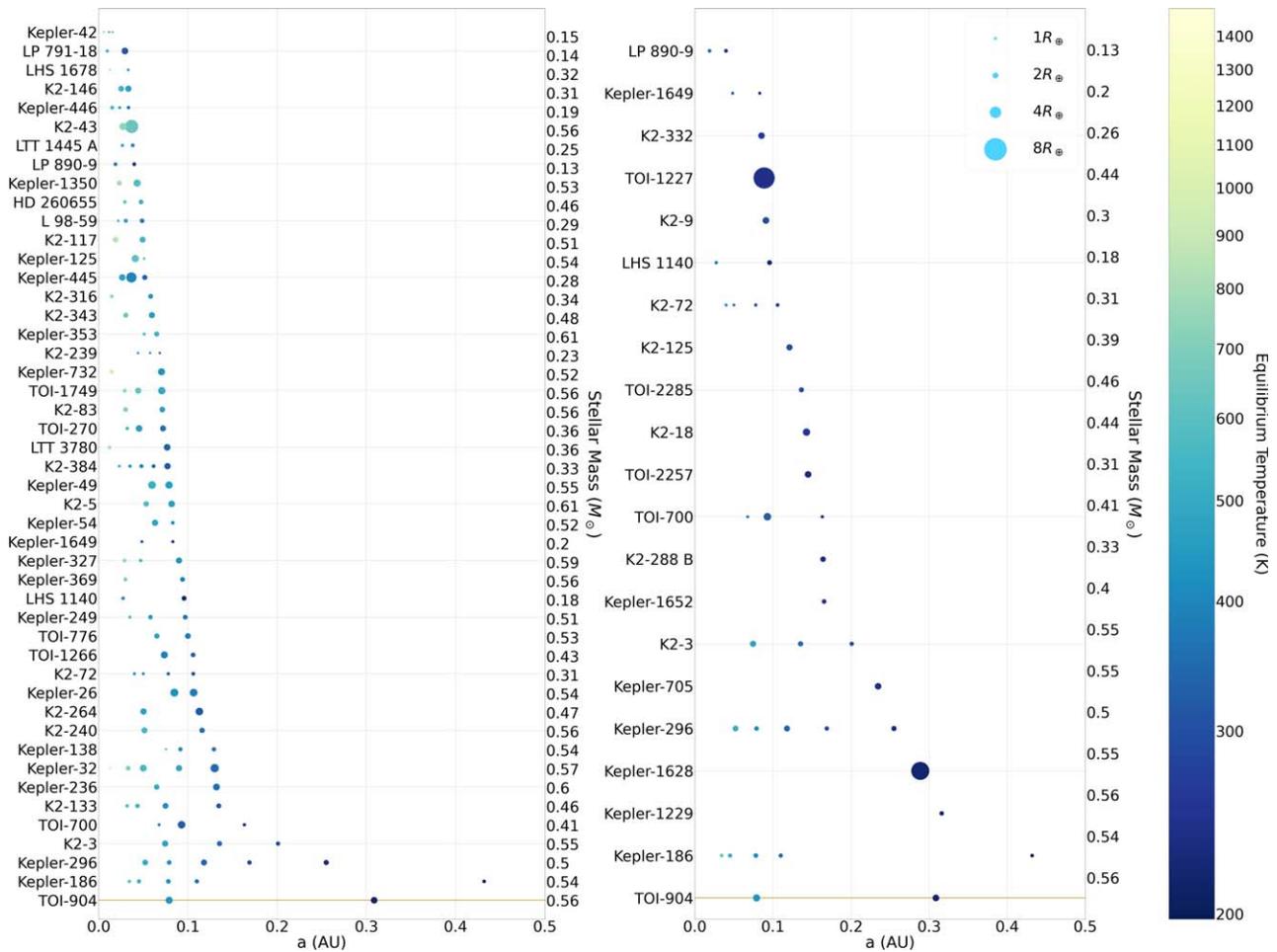

**Figure 3.** Plots of all known transiting multiplanet systems (left) and cold planets (right; $T_{eq} < 300$ K) around M dwarf stars. Each system is shown as a function of the system name (left y-axis), stellar mass (right y-axis), semimajor axis (x-axis) planet equilibrium temperature (color), and planet radius (point size). TOI-904 is shown at the bottom of each panel with a gold horizontal line.

We fit the LCO observations described in Section 2.2.1 of TOI-904 b using the `juliet` software library and our own `emcee` fit as described above and using the `LDTK` package to calculate the star's limb-darkening parameters in the SLOAN/SDSS z passband. We found that the transit depths for TESS were within $1\sigma$ of the fit we performed on the 2 minute SPOC data, as shown in Appendix A, Table 2.

## 4. Discussion

In this Letter, we have announced the discovery and validation of TOI-904 c, a cool sub-Neptune ($T_{eq} \approx 217.26 \pm 10.22$ K; $R_p = 2.167^{+0.130}_{-0.118} R_\oplus$; $P = 83.999 \pm 0.001$ days) that is the coldest planet orbiting an M dwarf star discovered by TESS to date. We also report on TOI-904 b, another sub-Neptune ($R_p = 2.426^{+0.163}_{-0.157} R_\oplus$; $P = 10.8772 \pm 0.0003$ days) located much closer to the early M dwarf host star.

M dwarfs are known to host a number of Earth- to Neptune-sized planets (see Figure 3). These planetary systems are generally very compact and tend to exist within ~0.2 au of their host star. By contrast, TOI-904 is an extended system with ~0.23 au between the two planets. While distant planets have been found around other M dwarfs, all of these planets were found either on the periphery of a compact system comprised of several other small planets (e.g., Kepler-186) or as the lone planet in their systems (e.g., Kepler-1628 and Kepler-1229). As this is already a multiplanet system, we searched for indications of any other planets orbiting TOI-904 to determine if this system belongs in the former category. One way in which we looked for nontransiting planets was by searching for transit-timing variations (TTVs) in the TESS and LCO observations of TOI-904 b and the TESS observations of TOI-904 c. We found the maximum TTV amplitude of a sine wave fit with periods from 0.1 to 100 days to be 5–10 minutes for the inner planet and 25–30 minutes for the outer planet (see transit times presented in Appendix B, Table 3). We found no current evidence of a nontransiting planet in the TTVs of the observed transits for TOI-904 b or c.

We also used the `dynamite` software library created by Dietrich & Apai (2020) to statistically calculate the potential periods and probabilities of unseen planets in multiplanet systems. `dynamite` calculates these probabilities by implementing a triple integral over the probability density function (PDF) of planet inclination, period, and radius based on the occurrence rates calculated by Mulders et al. (2018), assuming each variable is independent (Dietrich & Apai 2020). This PDF is then sampled using the Monte Carlo method before producing all dynamically stable results based on calculations using Dietrich & Apai's (2020) Equation (6) (see Figure 4). Our implementation of `dynamite` predicted four potential planets in the TOI-904 system with orbital periods of 6.74, 18.5, 50.7, and 139.05 days (see Figure 4). These additional planets may be detectable using extreme-precision radial velocity observations or potential TTVs in future observations of TOI-904 b and c, both of which will be reobserved by TESS in Sectors 65, 66, and 67.

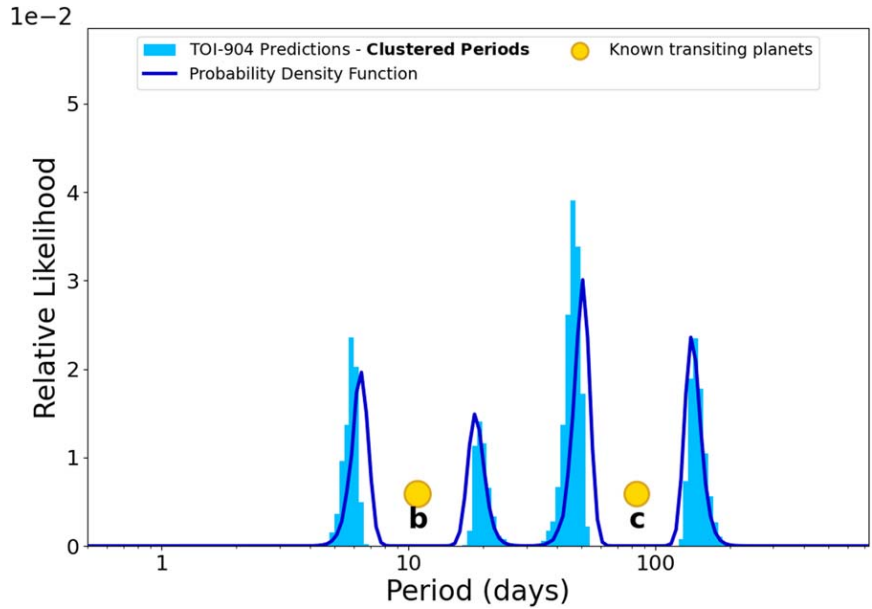

**Figure 4.** Created using `dynamite` (Dietrich & Apai 2020), the figure shows the relative likelihood of the most probable locations of unseen planets orbiting TOI-904 in log-period space. The yellow circles indicate the periods of the two known planets in the system, while the dark blue lines show the PDF calculated for this system, and the light blue histograms represent the stable Monte Carlo iterations sampled from the PDF. These scenarios were calculated using the `dynamite` Exoplanets Systems Simulator model (`syssim`; He et al. 2019). By incorporating the location of known planets and the stellar type, the `dynamite` software finds that additional nontransiting planets are most likely to have periods of 6.74, 18.5, 50.7, and 139.05 days.

Especially in the absence of another planet, the planets orbiting TOI-904 represent an interesting case study for exoplanet formation scenarios around low-mass stars. TOI-904 c is one of three known transiting M dwarf planets that orbit beyond 0.2 au and have $R_p > 1.8\,R_{\rm Earth}$, implying that they could have a gaseous envelope. The other two (Morton et al. 2016; Berger et al. 2018), however, are Kepler systems with $J_{\rm mag} > 12$, while TOI-904 has $J_{\rm mag} = 9.6$. This system thus provides an unprecedented opportunity to constrain planet formation at larger distances from M dwarfs by probing its atmosphere. Assuming a predicted mass of 6.1 $M_\oplus$ (based on the mass–radius relations of Chen & Kipping 2017), the transmission spectroscopy metric (TSM; Kempton et al. 2018) of TOI-904 c is 23. Thanks mainly to its higher equilibrium temperature, the TSM of TOI-904 b (assuming a mass of 6.9 $M_\oplus$) is 50. By comparison, the TSM values of Kepler-1628 b and Kepler-1229 b are 13 and 4, respectively.

Additional study of this system could resolve the currently ambiguous composition of the two planets. Simulations conducted by Burn et al. (2021) and Pan et al. (2022) of planet formation via core accretion around low-mass stars predict that planets with radii of $\sim2.3\,R_\oplus$ could have a range of densities spanning the ultradense sub-Neptune, water-world, and puffy sub-Neptune paradigms described by Luque & Pallé (2022). Both studies expect that in any case, planets of this size are likely to have some atmosphere, so it is not realistic to expect these planets to be rocky in composition. The mass limits we obtained using `radvel` on our CORALIE measurements rule out none of these scenarios, with $2\sigma$ upper limits of 174 and 244 $M_\oplus$ on the mass of TOI-904 b and c, respectively. With follow-up mass and atmospheric measurements, we can resolve this degeneracy for the outer planet and gain insight into formation of cold planets around low-mass stars. With the predicted masses mentioned above, we expect radial velocity semiamplitudes of 2.9 and 1.3 m s$^{-1}$ for TOI-904 b and c, respectively. Based on Luque & Pallé's (2022) analysis of M dwarf small planets, the three different regimes of planet density differ in scale by a factor of 2 (0.25, 0.5, and 1 $\rho_\oplus$ for puffy, water, and rocky planets, respectively). The masses for planets in each regime would similarly differ, resulting in detectable differences in the expected semiamplitudes of radial velocity measurements. We find that the three compositions would result in masses (semiamplitudes) of 3.3 $M_\oplus$ (1.4 m s$^{-1}$), 6.5 $M_\oplus$ (2.9 m s$^{-1}$), and 13.1 $M_\oplus$ (5.8 m s$^{-1}$), respectively, for the inner planet; for the outer planet, the masses (semiamplitudes) would be 2.6 $M_\oplus$ (1.1 m s$^{-1}$), 5.2 $M_\oplus$ (2.3 m s$^{-1}$), and 10.4 $M_\oplus$ (4.6 m s$^{-1}$), respectively. While a $5\sigma$ mass measurement (Batalha et al. 2019) could be achieved with HARPS (Mayor et al. 2003) or Magellan II/PFS (Crane et al. 2010; Teske et al. 2016), particularly for TOI-904 b, the mass precision needed to distinguish between different bulk compositions is realistically only within reach of VLT-ESPRESSO (Pepe et al. 2021). Indeed, ESPRESSO was used to measure a semiamplitude of 2.2 m s$^{-1}$ for the candidate planet LHS 1140d, which has a period of 79 days and orbits a star five times fainter than TOI-904 (Lillo-Box et al. 2020).

If TOI-904 c is an ultradense Neptune and thus most easily detectable for mass measurements, both planets may have formed in situ (Kennedy et al. 2006; Hansen & Murray 2012; Hansen 2015). This case would have resulted in TOI-904 c forming as a rocky, ultradense Neptune (similar to K2-110 b; Osborn et al. 2017) if the snow line of the M dwarf receded toward the star on planet formation timescales causing a late stage of mass accretion for planets located near the snow line (Kennedy et al. 2006). In situ formation seems unlikely for this system, however, as the larger inner planet, TOI-904 b, could not have reached its size at its current location due to the high temperature (>2000 K; Ali-Dib et al. 2020) of a forming protoplanet at this distance from the host star, precluding the accretion of any gaseous envelope. In situ formation also often predicts 4–10 rocky planets on coplanar orbits (Hansen & Murray 2012; Pan et al. 2022), for which there is currently no evidence in this system.

It is more probable that TOI-904 c migrated to its current location from a greater distance via type I migration (Burn et al. 2021; Luque & Pallé 2022; Pan et al. 2022). If the planet did form further in the disk, its core was likely formed of both rock and ice materials before the planet migrated inward. In this case, this planet's density could allow us to differentiate between a "water world" with ∼50% ice/rock ratios and little to no envelope (Burn et al. 2021; Luque & Pallé 2022) and a planet with a lower ice/rock ratio and a larger atmosphere (Pan et al. 2022). In this model, the inner planet's core could be made up of rocky materials or an ice/rock mixture. If the latter, Bitsch et al. (2019) stated that the hot inner water world would be considerable evidence for the planet migration model, and TOI-904 b's overall water content could inform the exact migration history of that planet. If, instead, planet b has a large gaseous envelope, its water content could still inform which formation path this system followed and where it acquired its envelope, either beyond the snow line (Burn et al. 2021) or after being trapped in the inner regions of the protoplanetary disk, where little water is available to accrete (Pan et al. 2022).

Type I migration is a very efficient process, however, and models of this theory predict very compact multiplanet systems, unlike the widely separated planets orbiting TOI-904. One possibility to resolve this discrepancy is that TOI-904 c formed at a significantly greater distance from the host star than TOI-904 b, which would explain why the inner planet migrated in so much further and why the mutual separation of the planets has persisted. This could also result from a lower disk density at greater distances, where type I migration could be triggered for smaller planets like TOI-904 c (Burn et al. 2021). This theory could be tested by determining if TOI-904 c has a larger C/O ratio than TOI-904 b, which would indicate that TOI-904 c may have formed beyond the methane ice line and that TOI-904 b did not.

Given that the planets are so similar in size, a final alternate theory could be that they followed the same formation path by forming sequentially, following the inside-out formation theory suggested by Chatterjee & Tan (2014). In this theory, pebbles form in outer regions of this disk before migrating inward and stalling at a high-pressure region (potentially at the water-ice line) before forming a planet at that location. The planet would then migrate inward and cause the high-pressure zone to move outward, allowing the process to repeat and form another planet. Pan et al. (2022) conducted the first simulations of this type of formation around M dwarfs and found that it is likely to result in systems of two to four sub-Neptune-sized planets located at greater distances and separations than type I migration would produce and thus more closely resembling the TOI-904 system. If both planets have a high water content, TOI-904 b's larger size could be explained by an inflated hydrosphere caused by its closer proximity to the star (Luque et al. 2019). If the compositions of these planets are similar, the TOI-904 system could strongly support this relatively new theory of planet formation. Whatever theory of formation dominates in these systems, through investigating the mass and atmospheric composition of both planets, we can determine whether they have twin compositions or are only sibling planets orbiting the same star.

## Acknowledgments

We thank the anonymous reviewer for the feedback and suggestions that have helped improve this paper. We would also like to acknowledge Rebekah Dawson at the Center for Exoplanets and Habitable Worlds at Pennsylvania State University for her advice in regards to the TTV analysis included in this paper.

Some of the data presented in this paper were obtained from the Mikulski Archive for Space Telescopes (MAST) at the Space Telescope Science Institute. The specific observations analyzed can be accessed via DOI:10.17909/73w3-v751. This research has made use of the NASA Exoplanet Archive, which is operated by the California Institute of Technology, under contract with the National Aeronautics and Space Administration under the Exoplanet Exploration Program. The data from the Hubble Tarantula Treasury Project (Sabbi 2016) may be obtained from MAST doi: 10.17909/T9RP4V. We acknowledge the use of public TESS data from pipelines at the TESS Science Office and the TESS Science Processing Operations Center. Resources supporting this work were provided by the NASA High-End Computing (HEC) Program through the NASA Advanced Supercomputing (NAS) Division at Ames Research Center for the production of the SPOC data products. Some of the observations in the paper made use of the high-resolution imaging instrument ZORRO obtained under Gemini LLP proposal No. GN/S-2021A-LP-105. ZORRO was funded by the NASA Exoplanet Exploration Program and built at the NASA Ames Research Center by Steve B. Howell, Nic Scott, Elliott P. Horch, and Emmett Quigley. ZORRO was mounted on the Gemini South telescope of the international Gemini Observatory, a program of NSF's OIR Lab, which is managed by the Association of Universities for Research in Astronomy (AURA) under a cooperative agreement with the National Science Foundation on behalf of the Gemini partnership: the National Science Foundation (United States), National Research Council (Canada), Agencia Nacional de Investigación y Desarrollo (Chile), Ministerio de Ciencia, Tecnología e Innovación (Argentina), Ministério da Ciência, Tecnologia, Inovações e Comunicações (Brazil), and Korea Astronomy and Space Science Institute (Republic of Korea). This work makes use of observations from the LCOGT network. Part of the LCOGT telescope time was granted by NOIRLab through the Mid-Scale Innovations Program (MSIP). MSIP is funded by NSF. This research has made use of the Exoplanet Follow-up Observation Program website (NExScI 2022), which is operated by the California Institute of Technology, under contract with the National Aeronautics and Space Administration under the Exoplanet Exploration Program. This work was based in part on observations obtained at the Southern Astrophysical Research (SOAR) telescope, which is a joint project of the Ministério da Ciência, Tecnologia e Inovações (MCTI/LNA) do Brasil, the US National Science Foundation's NOIRLab, the University of North Carolina at Chapel Hill (UNC), and Michigan State University (MSU). This work has made use of data from the European Space Agency (ESA) mission Gaia (https://www.cosmos.esa.int/gaia), processed by the Gaia Data Processing and Analysis Consortium (DPAC; https://www.cosmos.esa.int/web/gaia/dpac/consortium). Funding for the DPAC has been provided by national institutions, in particular the institutions participating in the Gaia Multilateral Agreement. This publication makes use of data products from the Two Micron All Sky Survey, which is a joint project of the University of Massachusetts and the Infrared Processing and Analysis Center/California Institute of Technology, funded by the National Aeronautics and Space Administration and the National Science Foundation. This publication makes use of data products from the Wide-field Infrared Survey Explorer, which is a joint project of the University of California, Los Angeles, and the Jet Propulsion Laboratory/California Institute of Technology, funded by the National Aeronautics and Space Administration.

D.D. acknowledges support from TESS Guest Investigator Program grant 80NSSC22K0185 and NASA Exoplanet Research Program grant 18-2XRP18_2-0136. The contributions of S.U., M.L., and H.O. have been carried out within the framework of the NCCR PlanetS supported by the Swiss National Science Foundation under grants 51NF40_182901 and 51NF40_205606. M.L. further acknowledges the support of the Swiss National Science Foundation under grant No. PCEFP2_194576.

*Facilities:* LCOGT, Gemini South/ZORRO, EULER/COR-ALIE, SOAR/HRCAM.

*Software:* AstroImageJ (Collins et al. 2017), TAPIR (Jensen 2013), astropy (Astropy Collaboration et al. 2013, 2018, 2022), triceratops (Giacalone & Dressing 2020), juliet (Espinoza et al. 2019), emcee (Foreman-Mackey et al. 2013), batman (Kreidberg 2015), wotan (Hippke et al. 2019), eleanor (Feinstein et al. 2019), dynesty (Speagle 2020), LDTK (Husser et al. 2013; Parviainen & Aigrain 2015), SpecMatch-Emp (Yee et al. 2017).

## Appendix A

Figures 5 and 6 show the transit models created using observations from the two-minute TESS-SPOC, 30-minute TESS-SPOC eleanor, and QLP pipeline observations for both TOI-904 b and c using both the individual and joint fits in the juliet library. Table 2 presents the transit model parameters for both planets based on observations from each pipeline along with the LCO observations of TOI-904 b (disccussed in Section 2.2.1) as fit using the juliet joint fit, the juliet individual fit, and the emcee/batman fit.

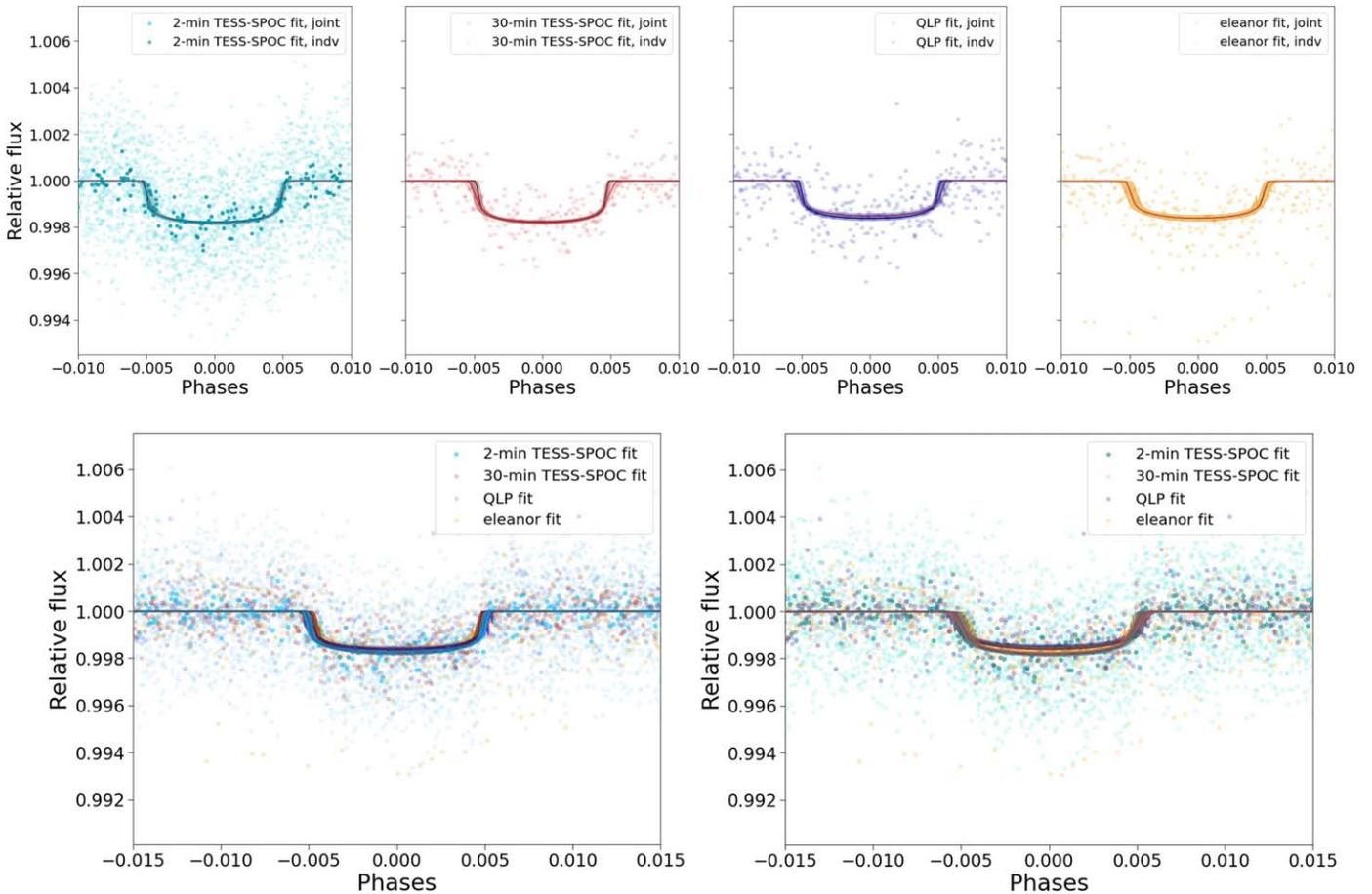

**Figure 5.** Graphic depictions of the juliet fits of TOI-904 b created using the different pipelines' data products. Top: data from each pipeline (lighter points, overlaid with darker points when binning was applied) with the joint and individual planet fits overplotted. Source of data products from left to right are the SPOC 2 minute, TESS-SPOC 30 minute, MIT Quicklook, and eleanor pipelines. Bottom left: comparison of each joint planet fit of TOI-904 b created from the different pipelines, with the SPOC 2 minute data in light blue, the 30 minute TESS-SPOC data in red, the QLP data in dark blue, and the eleanor data in yellow. Bottom right: same as bottom left panel but for the individual planet fits of TOI-904 b.

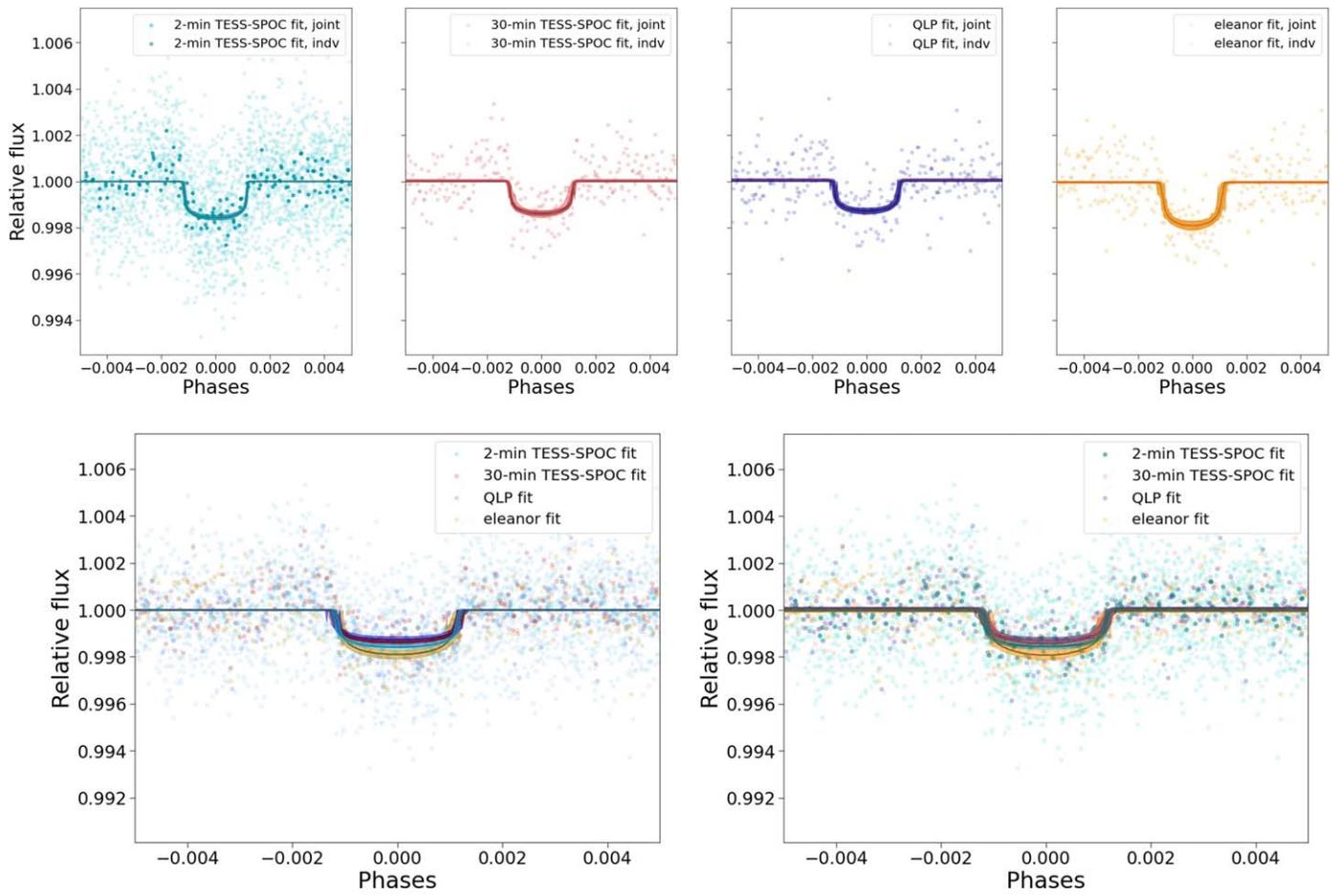

**Figure 6.** Same as Figure 5 but fitting for TOI-904 c.

## Appendix B

Table 3 presents the mid-transit times obsered by TESS for both TOI-904 b and c when the host star was observed in TESS Sectors 12, 13, 27, 38, 39, and 61.

**Table 3**
The Transit Times for All TESS Transits of TOI-904 b and c as They Were Observed in Sectors 12, 13, 27, 38, 39, and 61

| Planet | Transit Time (BJD-TDB) |
|---|---|
| TOI-904 b | $1637.847^{+0.005}_{-0.005}$ |
|  | $1648.716^{+0.006}_{-0.006}$ |
|  | $1659.599^{+0.009}_{-0.006}$ |
|  | $1670.478^{+0.009}_{-0.003}$ |
|  | $1681.374^{+0.010}_{-0.015}$ |
|  | $2040.301^{+0.003}_{-0.004}$ |
|  | $2051.181^{+0.003}_{-0.004}$ |
|  | $2344.870^{+0.003}_{-0.005}$ |
|  | $2355.746^{+0.004}_{-0.004}$ |
|  | $2366.630^{+0.006}_{-0.008}$ |
|  | $2377.510^{+0.004}_{-0.004}$ |
|  | $2388.369^{+0.006}_{-0.007}$ |
|  | $2964.869^{+0.005}_{-0.006}$ |
|  | $2986.621^{+0.005}_{-0.010}$ |
| TOI-904 c | $1630.353^{+0.003}_{-0.004}$ |
|  | $2050.349^{+0.005}_{-0.006}$ |
|  | $2386.358^{+0.006}_{-0.011}$ |
|  | $2974.345^{+0.007}_{-0.009}$ |

**Note**: Two transits of TOI-904 b were excluded due to poor quality data near 2333.999 and 2975.741.

## ORCID iDs

Mallory Harris https://orcid.org/0000-0002-3721-1683
Diana Dragomir https://orcid.org/0000-0003-2313-467X
Ismael Mireles https://orcid.org/0000-0002-4510-2268
Karen A. Collins https://orcid.org/0000-0001-6588-9574
Steve B. Howell https://orcid.org/0000-0002-2532-2853
Keivan G. Stassun https://orcid.org/0000-0002-3481-9052
George Zhou https://orcid.org/0000-0002-4891-3517
Carl Ziegler https://orcid.org/0000-0002-0619-7639
François Bouchy https://orcid.org/0000-0002-7613-393X
César Briceño https://orcid.org/0000-0001-7124-4094
David Charbonneau https://orcid.org/0000-0002-9003-484X
Kevin I. Collins https://orcid.org/0000-0003-2781-3207
Natalia M. Guerrero https://orcid.org/0000-0002-5169-9427
Jon M. Jenkins https://orcid.org/0000-0002-4715-9460
Eric L. N. Jensen https://orcid.org/0000-0002-4625-7333
Martti H. K. Kristiansen https://orcid.org/0000-0002-2607-138X
Nicholas Law https://orcid.org/0000-0001-9380-6457
Monika Lendl https://orcid.org/0000-0001-9699-1459
Andrew W. Mann https://orcid.org/0000-0003-3654-1602
Hugh P. Osborn https://orcid.org/0000-0002-4047-4724
Samuel N. Quinn https://orcid.org/0000-0002-8964-8377
George R. Ricker https://orcid.org/0000-0003-2058-6662
Richard P. Schwarz https://orcid.org/0000-0001-8227-1020
Sara Seager https://orcid.org/0000-0002-6892-6948
Eric B. Ting https://orcid.org/0000-0002-8219-9505
Roland Vanderspek https://orcid.org/0000-0001-6763-6562
David Watanabe https://orcid.org/0000-0002-3555-8464
Joshua N. Winn https://orcid.org/0000-0002-4265-047X